\documentclass[twocolumn,showpacs,preprintnumbers,amsmath,amssymb,prb]{revtex4-1}
\usepackage{graphicx}
\usepackage{dcolumn}
\usepackage{bm}

\begin{document}

\title{Signatures of spin-glass behavior in the induced magnetic moment system PrRuSi$_{3}$}
\author{V. K. Anand}
\altaffiliation{vivekkranand@gmail.com, Present address: Ames Laboratory, Department of Physics and Astronomy, Iowa State University, Ames, Iowa 50011, USA.}
\author{D. T. Adroja}
\email{devashibhai.adroja@stfc.ac.uk}
\author{A. D. Hillier}
\author{J. Taylor}
\affiliation{ISIS Facility, Rutherford Appleton Laboratory, Chilton, Didcot Oxon, OX11 0QX, UK}
\author{G. Andr\'e}
\affiliation{Laboratoire Leon Brillouin (CEA-CNRS), CEA/Saclay, 91191 Gif-sur-Yvette, France}

\date{\today}

\begin{abstract}
We have investigated the magnetic and transport properties of a ternary intermetallic compound PrRuSi$_{3}$ using dc magnetization, ac susceptibility, specific heat, electrical resistivity, neutron diffraction, inelastic neutron scattering and $\mu$SR measurements. The magnetic susceptibility and specific heat data reveal the signatures of spin-glass behavior in PrRuSi$_{3}$ with a freezing temperature of 9.8 K. At low magnetic fields, we observe two sharp anomalies (at 4.9 and 8.6 K) in magnetic susceptibility data. In contrast, the specific heat data show only a broad Schottky-type anomaly centered around 10 K. However, $\mu$SR reveals very low frequency coherent oscillations at 1.8 K with an onset of fast relaxation below 12 K indicating a long range magnetically ordered ground state with very small moment. On the other hand, no magnetic Bragg peaks are observed in low temperature neutron diffraction data at 1.8 K. These two contradictory ground states, spin-glass versus magnetic order, can be explained if the spin-glass behavior in PrRuSi$_3$ is considered due to the dynamic fluctuations of the crystal field levels as has been proposed for spin-glass behavior in PrAu$_{2}$Si$_{2}$. Two sharp inelastic excitations near 2.4 meV and 14.7 meV are observed in the inelastic neutron scattering (INS) spectra between 4 K and 50 K. Further, exchange coupling $J_{ex}$ obtained from the analysis of INS data with CEF model provides evidence for the spontaneously induced magnetic order with a CEF-split singlet ($\Gamma_{t4}$) ground state. However, the exchange coupling seems to be close to the critical value for the induced moment magnetism, therefore we tend to believe that the dynamic fluctuations between the ground state singlet and excited doublet CEF levels is responsible for spin-glass behavior in PrRuSi$_3$.
\end{abstract}


\maketitle

\section{Introduction}

Frustrated systems can present a wide range of physical phenomena due to more configuration options than the ordered systems, and, as such have been paid particular attention in recent past and these systems have emerged as a topic of contemporary research in condensed matter physics. \cite{1,2,3,4,5,6,7,8,9,10,11} Spin-glass systems are characterized by a random frozen spin orientation below a characteristic freezing temperature and present a nice example of frustrated magnetism. In contrast to a unique ground state found in conventional collective systems, spin-glass systems possess a multitude of possible disordered ground states and freeze into one of them. Investigations into spin-glass systems have shown that frustration and randomness are the key ingredients for spin-glass behavior. The crystallographic disorder in a magnetic system may cause frustration of the magnetic moments leading to spin-glass behavior. However, for a well-ordered crystal structure there is no obvious source of frustration to magnetic moments and, therefore, one would not expect spin-glass behavior in crystallographically ordered compounds (except for the specific class of geometrically frustrated lattice, i.e. due to antiferromagnetic coupling among the magnetic moments in a triangular lattice). For this reason research into spin-glass behavior in magnetic systems has been generally focused on geometrically frustrated kagome (two dimensional) and pyrochlore (three-dimensional) lattices. \cite{12}

However, the recent observations of spin-glass behavior in the well-ordered stoichiometric intermetallic compounds URh$_{2}$Ge$_{2}$ \cite{13} and PrAu$_{2}$Si$_{2}$ \cite{14} have brought further scope and insight into the mechanism of spin-glass behavior.  The spin-glass behavior in URh$_{2}$Ge$_{2}$ was explained in terms of the site disorder on the rhodium and germanium sublattices and, extended annealing was found to remove the disorder resulting in an antiferromagnetically ordered state. \cite{15} In contrast, the spin-glass behavior in PrAu$_{2}$Si$_{2}$ was found to persist despite extended annealing. This suggests that the spin-glass behavior in PrAu$_{2}$Si$_{2}$ is not driven by the crystallographic disorder. A novel mechanism due to the dynamic fluctuations of the crystal field levels has been proposed to be responsible for the frustration of the magnetic moments in PrAu$_{2}$Si$_{2}$.\cite{16} In this paper we present another example of well-ordered stoichiometric intermetallic system PrRuSi$_{3}$ that exhibit the signatures of spin-glass behavior most likely through the dynamic fluctuations of the crystal field levels. Further, the spin-glass state in PrRuSi$_{3}$ is found to compete with the spontaneously induced magnetic order with a CEF singlet ground state. Its Ce-analog, CeRuSi$_{3}$ is reported to exhibit no magnetic order down to 2 K both for polycrystalline and single crystal samples. \cite{17,18} A broad maximum is observed in the magnetic susceptibility of CeRuSi$_{3}$ near 150 K which indicates the presence of strong hybridization between the 4$f$ and conduction electrons.

\section{Experimental}

Polycrystalline samples of PrRuSi$_{3}$ and its nonmagnetic analog LaRuSi$_{3}$ were prepared using standard arc melting technique on a water cooled copper hearth under the titanium gettered inert argon atmosphere using the high purity elements (Pr,La:  99.9\%, Ru: 99.99\%, Si: 99.999\%) in stoichiometric ratio. During the melting process the samples were flipped and remelted several times to achieve homogeneity. Further, to improve homogeneity and phase formation, and reduce disorder (if any present) the as-cast samples were wrapped in tantalum foil and annealed for a week at 900~$^{o}$C under the dynamic vacuum. The crystal structure and phase purity were checked by the powder X-ray diffraction (XRD) and scanning electron microscopy (SEM). The stoichiometry was checked by the energy dispersive X-ray (EDAX) composition analysis. A commercial SQUID magnetometer (MPMS, Quantum-Design) was employed for magnetization measurements. Specific heat was measured by the relaxation method in a physical properties measurement system (PPMS, Quantum-Design).  Electrical resistivity was measured by standard four-probe ac technique using the PPMS. The $\mu$SR measurements were performed in longitudinal geometry using the MuSR spectrometer at ISIS facility of the Rutherford Appleton Laboratory, Didcot, UK.  Since silver gives a non-relaxing muon signal we used a silver holder (purity 4N) for mounting the sample. The stray fields at the sample position were cancelled to within 1 $\mu$T by using correction coils. The neutron scattering experiment was performed on the powdered PrRuSi$_{3}$ sample in the temperature range 4 to 50 K using the MARI time of flight spectrometer at ISIS. The powdered sample was mounted by wrapping the sample in a thin Al-foil inside a thin walled Al-can. The low temperature was achieved using He-exchange gas with a top-loading closed cycle refrigerator. The data were collected using the neutron incident energy E$_{i}$ = 6, 20 and 40 meV. The neutron diffraction measurements were performed on the powdered sample using the G4.1 diffractometer at LLB, Saclay. The neutron diffraction data were collected at 1.8 and 25 K using the neutrons of wavelength 2.423 {\AA}.

\section{Results and discussion}

\begin{figure}
\includegraphics[width=9.0cm, keepaspectratio]{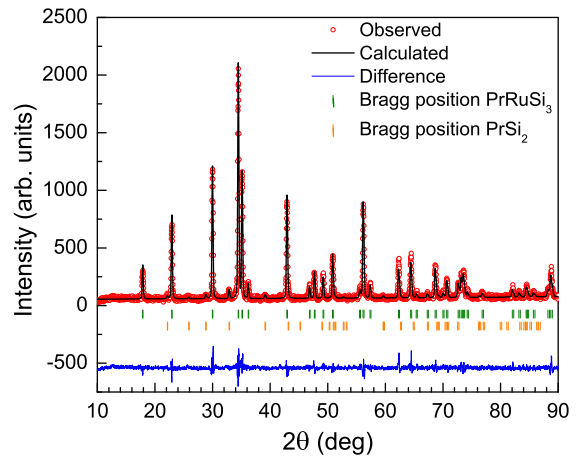}
\caption{\label{fig1} (colour online) Powder X-ray diffraction pattern of PrRuSi$_{3}$ recorded at room temperature. The solid line through the experimental points is the two-phase Rietveld refinement profile calculated for BaNiSn$_{3}$-type tetragonal (space group \textit{I4 mm}) and ThSi$_2$-type tetragonal (space group {\it I4$_1$/amd}) structural model. The vertical bars indicate the Bragg positions of both the phases. The lowermost curve represents the difference between the experimental data and calculated results.}
\end{figure}

\begin{table}
\caption{\label{tab:table1} Crystallographic parameters for PrRuSi$_{3}$ determined from the full structure refinement of powder X-ray diffraction data using Fullprof software.}
\begin{ruledtabular}
\begin{tabular}{llcccc}

Structure & BaNiSn$_{3}$-type tetragonal\\

Space group & \textit{I4 mm} (No. 107)\\

Crystal parameters\\

  \hspace{1.0 cm}    $a$        &  4.213(1) {\AA} \\

  \hspace{1.0 cm}    $c$       &   9.923(1) {\AA} \\

  \hspace{1.0 cm}  $V_{cell}$  &   176.01(1) {\AA}$^{3}$ \\

Atomic Coordinates \\

   \hspace{0.75 cm} Atom (site) & x  \hspace{1.0 cm} y   \hspace{1.0 cm} z \\

   \hspace{1.0 cm}    Pr (2a)      & 0 \hspace{1.0 cm} 0   \hspace{1.0 cm} 0.5778(13) \\

   \hspace{1.0 cm}    Ru (2a)      & 0   \hspace{1.0 cm} 0   \hspace{1.0 cm} 0.2343(14) \\

   \hspace{1.0 cm}    Si1 (2a)      & 0  \hspace{1.0 cm} 0   \hspace{1.0 cm} 0 \\

   \hspace{1.0 cm}    Si2 (4b)      & 0  \hspace{1.0 cm} 0.5 \hspace{0.75 cm} 0.3439(18) \\

\end{tabular}
\end{ruledtabular}
\end{table}

The analysis of powder X-ray diffraction data reveals that PrRuSi$_{3}$ crystallizes in the BaNiSn$_{3}$-type tetragonal structure (space group \textit{I4 mm}). Fig.~1 shows the XRD pattern together with the two-phase Rietveld refinement profile fit, which indicates a well crystallographic ordered structure and the single phase nature of sample, the impurity phase identified as PrSi$_2$ is very small (1.54 \% in volume fraction, equivalent to 2.19 \% mole fraction).  The lattice parameters were found to be $a$ = 4.213(1) {\AA} and $c$ = 9.923(1) {\AA}. The crystallographic parameters obtained from the Rietveld refinement of PrRuSi$_{3}$ are listed in Table~I. LaRuSi$_{3}$ also crystallizes in similar BaNiSn$_{3}$-type tetragonal structure (space group \textit{I4 mm}) with lattice parameters $a$ = 4.263(1) {\AA} and $c$ = 9.944(1) {\AA}.  For the best fits the least squares refinement gave the value of  $\chi^{2}$ = 1.84 and 1.47 respectively for PrRuSi$_{3}$ and LaRuSi$_{3}$. The single phase nature of PrRuSi$_{3}$ and LaRuSi$_{3}$ samples were further confirmed by the high resolution SEM (scanning electron microscopy) images, and the EDAX composition analysis confirmed the desired 1:1:3 stoichiometry.

The impurity phase PrSi$_2$ which forms in ThSi$_2$-type tetragonal (space group {\it I4$_1$/amd}) at room temperature is reported to exhibit ferromagnetic order below 11 K followed by another transition near 7 K, and undergoes a structural phase transition to GdSi$_2$-type orthorhombic structure (space group {\it Imma}) below 153 K. \cite{19,20,21} Coincidentally the dc magnetic susceptibility of our compound PrRuSi$_3$ also exhibits two sharp anomalies at 8.6 K and 4.9 K which are very close to what have been observed in PrSi$_2$. However, we do not observe any signature of ferromagnetic order in low temperature neutron diffraction data which suggests that the effect of impurity phase PrSi$_2$, if any, is very small. In fact from a systematic study of ac and dc magnetic susceptibility, specific heat, electrical resistivity, inelastic neutron scattering and $\mu$SR data we have found evidence of an induced moment spin-glass type behavior in this compound. The reproducibility of results have been checked by the magnetic susceptibility and specific heat measurements on two different batches of samples (having different impurity phases) and we believe that the small impurity content does not influence the results and conclusions derived on the bulk properties of PrRuSi$_3$ which we present in this paper.

\begin{figure}
\includegraphics[width=8.5cm, keepaspectratio]{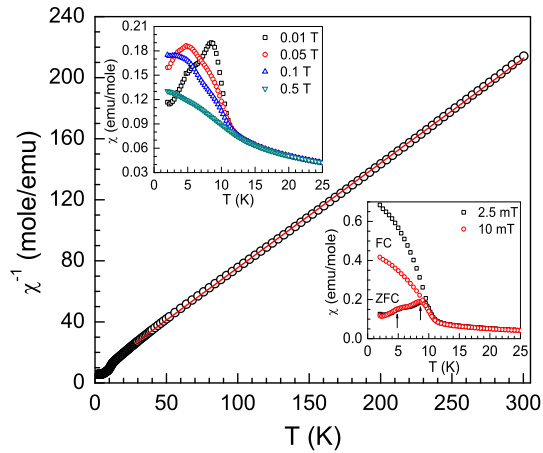}
\caption{\label{fig2} (colour online) The zero field cooled dc magnetic susceptibility $\chi(T)$ of PrRuSi$_{3}$ as a function of temperature in the temperature range 2 -- 300 K measured in a field of 0.1 T. The solid line represents the fit to Curie-Weiss behavior. The upper inset shows the expanded view of low temperature susceptibility measured at different applied magnetic fields. The lower inset shows the zero field cooled (ZFC) and field cooled (FC) susceptibility data at 2.5 and 10 mT.}
\end{figure}

\begin{figure}
\includegraphics[width=8.5cm, keepaspectratio]{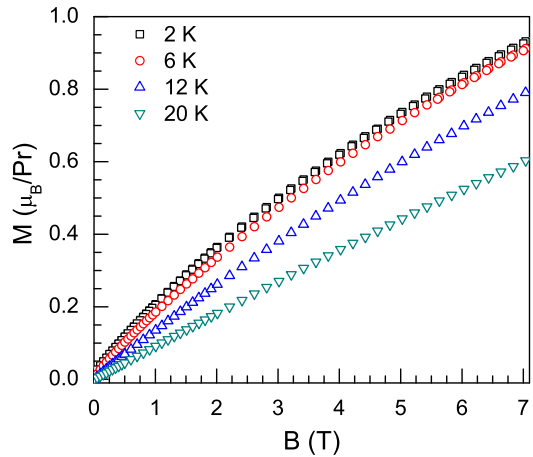}
\caption{\label{fig3} (colour online) The dc isothermal magnetization $M(B)$ as a function of magnetic field measured at constant temperatures of 2, 6, 12 and 20  K.}
\end{figure}

Results obtained from the dc magnetization measurements are shown in Fig.~2 and Fig.~3. At low fields (e.g., at 0.01 T) the low temperature magnetic susceptibility data clearly show two well defined anomalies at 8.6 K and 4.9 K (insets in Fig. 2). While with an increase in applied magnetic field the 4.9 K anomaly becomes more pronounced, the anomaly at 8.6 K gets smoothened, however, no change is observed in the temperature at which this anomaly occurs. The nature of these transitions at 4.9 K and 8.6 K is not clear from the temperature dependent dc magnetic susceptibility data, however the ac susceptibility and specific heat measurements as discussed later hint for a signature of spin-glass type behavior in this compound. Further, the behavior of magnetic anomaly at 4.9 K suggests that an increase in the magnetic field causes the spin alignment and the 4.9 K transition might be related to spin reorientation. On further increase in applied magnetic field, at an applied field of 0.5 T both the anomalies in the susceptibility are hardly detectable and susceptibility tends to saturate. In the lower inset of Fig.~2 we have plotted the zero field cooled (ZFC) and field cooled (FC) dc susceptibility data as a function of temperature at an applied field of 2.5 mT and 10 mT. The splitting of FC and ZFC susceptibility data can be taken as the first indication for the spin-glass behavior in this compound. The splitting of FC and ZFC susceptibility sets in at $T$ = 9.8 K defining the quasi-static freezing temperature ($T_{f}$) for the spin-glass transition. The high temperature magnetic susceptibility data are consistent with the Curie-Weiss behavior, $\chi(T) = C/(T-\theta_{p}$). From the fit to the inverse susceptibility data above 50 K (solid line in Fig.~2), we obtain an effective moment $\mu_{eff}$ = 3.41 $\mu_{B}$ and Curie-Weiss temperature $\theta_{p}$ = $-$10.0 K. The value of effective moment obtained is very close to the theoretically expected value of 3.58 $\mu_{B}$ for Pr$^{3+}$ free ions. Fig.~3 shows the isothermal magnetization data as a function of magnetic field measured at constant temperatures of 2, 6, 12 and 20 K. The isotherms exhibit almost a linear field dependence and attain a value of $\sim$ 0.9 $\mu_{B}$ at 7 T at 2 K which is very small compared to the theoretical saturation magnetization of 3.2 $\mu_{B}$, which indicates the presence of the crystal field effect that will split the 4$f$ multiplet ground state of the Pr-ion ($J=4$).

\begin{figure}
\includegraphics[width=8.5cm, keepaspectratio]{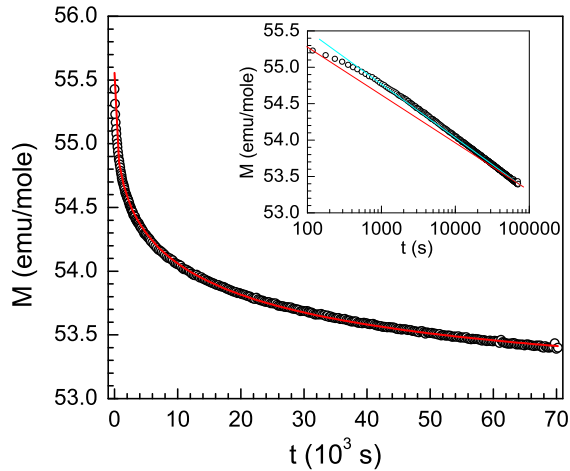}
\caption{\label{fig4} (colour online) The time dependence of thermo-remnant magnetization (TRM), $M(t)$ recorded at 2 K after switching off the cooling magnetic field of 0.05 T. The solid line is the fit to the superposition of a stretched exponential and a constant term. The inset shows the semi-log plot of the thermo-remnant magnetization. The solid lines in the inset are the guide to eyes.}
\end{figure}

In Fig.~4 we have shown the time dependence of the thermo-remnant magnetization (TRM), $M(t)$ which we recorded at 2 K, the sample was cooled in a magnetic field of 0.05 T from 50 K (well above $T_{f}$) to 2 K and the field-cooled isothermal remanence magnetization was measured after switching off the magnetic field. It is interesting to note that the magnetization does not become zero immediately after switching off the applied field as one expects for a normal antiferromagnetic order system, but it decays slowly with time as observed for many spin-glass systems. \cite{1} The non zero value of the magnetization after 70000 s, indicates frozen spin dynamics which transforms from one spin configuration to another spin configuration with time.  Many models have been proposed to describe the time dependence of magnetization relaxation in spin-glasses,\cite{22,23,24,25,26} the logarithmic relaxation decay, $M(t) = M_{0} - S log(t)$ and the stretched exponential decay $M(t) = M_{0} exp[-(t/\tau)^{1-n}]$ are the most common in literature. The inset of Fig.~4 shows the semi-log plot of TRM data, the data significantly deviates from the ideal linear behavior, probably due to the aging time effects. \cite{25} This suggests that a logarithmic relaxation decay is not a good choice for our data and the relaxation of thermo-remnant magnetization in PrRuSi$_{3}$ is much slower. A superposition of a stretched exponential and a constant term is found to represent the experimentally observed TRM data very well,
\[
M = M_{0} + M_{1} exp[-(t/\tau)^{1-n}]
\]
\noindent where the additional constant term is interpreted as the longitudinal spontaneous magnetization coexisting with the frozen transverse spin component. \cite{27} The solid line in Fig.~4 represents the fit to this expression, the best fit parameters are $M_{0}$ = 53.03 emu/mole, $M_{1}$ = 2.53 emu/mole, mean relaxation time $\tau$ = 13357 s and $n$ = 0.62.

\begin{figure}
\includegraphics[width=8.5cm, keepaspectratio]{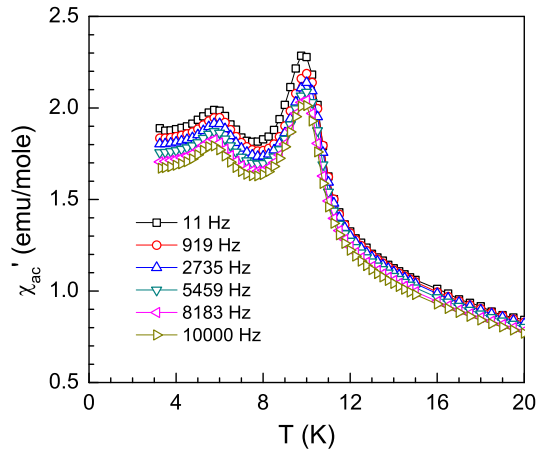}
\caption{\label{fig5} (colour online) The temperature dependence of the real part of ac magnetic susceptibility of PrRuSi$_{3}$ measured at different frequencies.}
\end{figure}

\begin{figure}
\includegraphics[width=8.5cm, keepaspectratio]{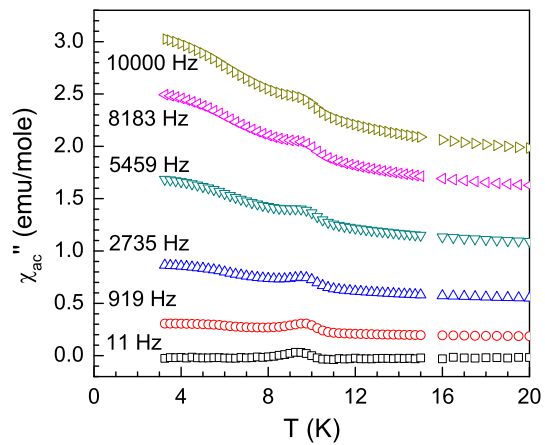}
\caption{\label{fig6} (colour online) The temperature dependence of the imaginary part of ac magnetic susceptibility of PrRuSi$_{3}$ measured at different frequencies.}
\end{figure}

In order to get further information about the magnetic behavior of PrRuSi$_{3}$ we also measured the ac magnetic susceptibility at different frequencies. The real and imaginary parts of ac magnetic susceptibility are shown in Fig.~5 and Fig.~6, respectively. At 11 Hz the real part of ac susceptibility $\chi_{ac}^{\prime}$ shows two sharp peaks at 5.7 K and 9.8 K. While there is no change in the peak position of the 5.7 K anomaly with frequency up to 10 kHz, the 9.8 K (11 Hz) peak shifts to 10.0 K at 919 Hz and remains unchanged up to 10 kHz. This frequency dependence is reminiscent of a spin glass. The imaginary component of ac susceptibility $\chi_{ac}^{\prime\prime}$ shows only one peak near 9.8 K which is independent of frequency and there is almost no noticeable anomaly (except a weak change in slope) near 5.7 K. The maximum of $\chi_{ac}^{\prime\prime}$ can be taken as the freezing temperature ($T_{f}$) of spin-glass behavior. However, the extremely weak (almost negligible) frequency dependence of ac susceptibility maximum suggests that the PrRuSi$_{3}$ is a non-canonical spin-glass system. For a canonical spin-glass system one expects an increase in transition temperature with increasing frequency, characterized by $\delta T_f = \Delta T_f/T_f \Delta log f$, for most of the canonical spin-glass system $\delta T_f$ is found to lie in between 0.0045 and 0.06. \cite{1} The 5.7 K anomaly in $\chi_{ac}^{\prime}$ can be attributed to the transition to a further frozen spin-glass state.

\begin{figure}
\includegraphics[width=8.5cm, keepaspectratio]{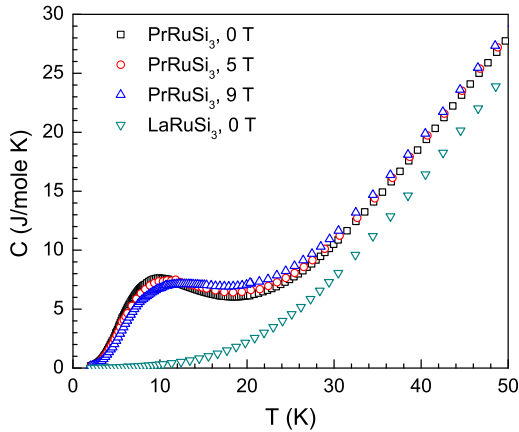}
\caption{\label{fig7} (colour online) The temperature dependence of specific heat data of LaRuSi$_{3}$ and PrRuSi$_{3}$ measured in different magnetic fields.}
\end{figure}

The specific heat data of PrRuSi$_{3}$ and LaRuSi$_{3}$ are shown in Fig.~7. The specific heat of PrRuSi$_{3}$ shows a very broad anomaly with a maximum near 10 K which we believe is due to crystal field effects. We do not see any sharp anomaly near 4.9 K or 8.6 K which would suggest a magnetic phase transition corresponding to the peaks in dc magnetic susceptibility data. With the application of magnetic field, the specific heat anomaly becomes broader and the position of maximum increases in temperature. The fact that the specific heat anomaly persists up to the investigated field of 9 T clearly rules out the possibility of the anomaly being related to a magnetic phase transition. The absence of the magnetic phase transition in specific heat is consistent with the spin-glass type behavior. The specific heat data of LaRuSi$_{3}$ does not show any anomaly down to 2 K and the low temperature data have a temperature dependence of $ C = \gamma T + \beta T^{3}$ with the Sommerfeld coefficient $\gamma$ $\sim$ 5 mJ/mole~K$^{2}$. In contrast, the specific heat of PrRuSi$_{3}$ does not obey the $ C = \gamma T + \beta T^{3}$ temperature dependence, the Sommerfeld coefficient $\gamma$ of PrRuSi$_{3}$ seems enhanced due to the crystal field effect with a C/T value of $\sim$ 93 mJ/mole~K$^{2}$ at 2 K. An extrapolation of low temperature specific heat data gives a rough estimate of $\gamma$ $\sim$ 58 mJ/mole K$^{2}$ for PrRuSi$_{3}$.

The magnetic contributions to the specific heat and entropy of PrRuSi$_{3}$ are shown in Fig.~8. The magnetic contribution to specific heat was obtained by subtracting the lattice contribution from the specific heat of PrRuSi$_{3}$ which we took roughly equal to the specific heat of LaRuSi$_{3}$. The magnetic entropy was obtained by integrating the $C_{mag} /T$ {\it vs.} $T$ plot. The broad Schottky-type anomaly observed in $C_{mag}$ is reasonably reproduced with the crystal field analysis. The solid line in Fig.~8 represents the fit to $C_{mag}$ data with the crystal field scheme obtained from the inelastic neutron scattering data discussed later. The ground state is a singlet and the first excited state is also a singlet at $\sim$  22 K, the second excited state being a doublet at $\sim$ 28 K. The temperature dependence of magnetic entropy is consistent with the singlet ground state and attains a value of $R ln 2$ near 12 K and $R ln 3$ near 22 K which supports the proposed crystal field scheme (the excess entropy being the contribution from higher excited states). The enhancement of Sommerfeld coefficient $\gamma$ can thus be understood in terms of excitonic mass enhancement due to low lying crystal field excitations as suggested in references 28 and 29 for a system with CEF-split singlet ground state.

\begin{figure}
\includegraphics[width=8.5cm, keepaspectratio]{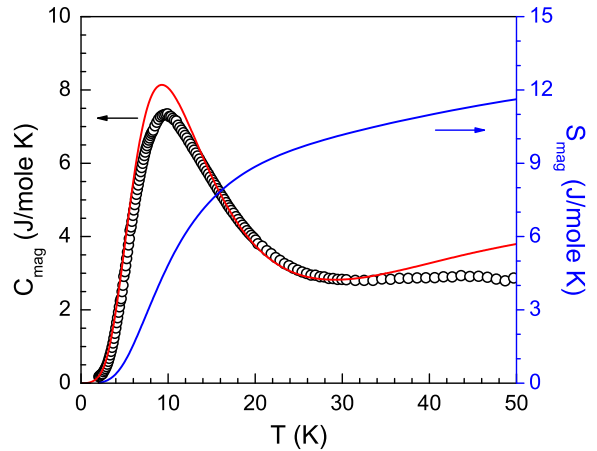}
\caption{\label{fig8} (colour online) The temperature dependence of magnetic contribution to the specific heat and entropy of PrRuSi$_{3}$ at zero field. The solid line shows the fit based on the crystal field model obtained from the inelastic neutron scattering data.}
\end{figure}

\begin{figure}
\includegraphics[width=8.5cm, keepaspectratio]{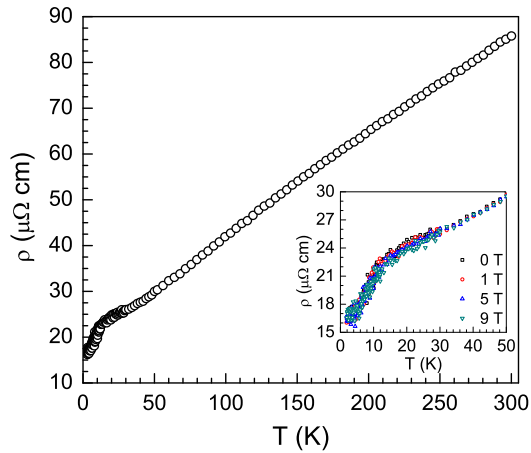}
\caption{\label{fig9} (colour online) The temperature dependence of electrical resistivity of PrRuSi$_{3}$ measured in zero magnetic field. The inset shows the expanded view below 50 K for different applied magnetic fields.}
\end{figure}

Figure 9 shows the electrical resistivity data of PrRuSi$_{3}$ measured in different applied magnetic fields. Above 50 K the electrical resistivity  exhibit metallic behavior and almost a linear temperature dependence with a residual resistivity of $\sim$  16 $\mu\Omega$~cm (at 1.85 K) and residual resistivity ratio of $\sim$ 5. The low temperature resistivity data show a broad curvature with a maximum near 15 K which can be attributed to the crystal field effect as revealed by the specific heat data discussed above. An extremely weak effect of magnetic field is observed on the electrical resistivity anomaly up to the investigated field of 9 T.

\begin{figure}
\includegraphics[width=8.5cm, keepaspectratio]{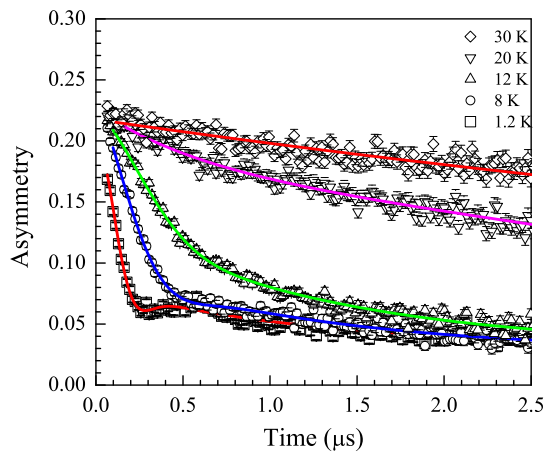}
\caption{\label{fig10} (colour online) The $\mu$SR spectra of PrRuSi$_{3}$ collected at various temperatures ($\square$ 1.2 K, $\circ$ 8 K, $\triangle$ 12 K, $\triangledown$ 20 K, $\diamond$ 30 K). The solid lines are the fits to the data as discussed in the text.}
\end{figure}

\begin{figure}
\includegraphics[width=8.5cm, keepaspectratio]{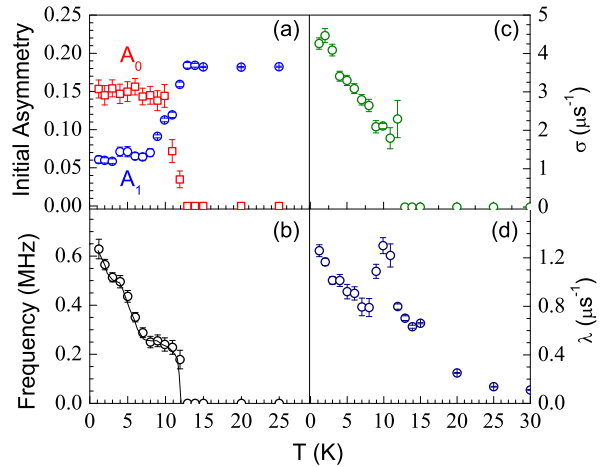}
\caption{\label{fig11} (colour online) The temperature dependence of (a) the initial asymmetries $A_{0}$ and $A_{1}$, (b) the precession frequency $\omega$, (c) the depolarization rate $\sigma$, and (d) the depolarization rate $\lambda$. The solid line in (b) is a guide to the eye.}
\end{figure}

In order to further probe the nature of the ground state of PrRuSi$_{3}$, spin-glass behavior versus long range antiferromagnetic order, we have carried out zero-field $\mu$SR study.  The $\mu$SR spectra were collected while warming the sample in zero field. Fig.~10 shows $\mu$SR spectra of PrRuSi$_{3}$ at various temperatures in the temperature range 1.2 -- 30 K. A marked change is observed in the muon depolarization rate for the temperatures above and below 12 K (Fig.~10), the temperature close to 9 - 10 K at which we have observed the anomaly in the dc and ac magnetic susceptibility. Above 12 K, the $\mu$SR spectra can be described by a simple exponential decay, indicating that the muons are sensing paramagnetic fluctuations. However, below 12 K the $\mu$SR spectra are best described by a heavily damped oscillating function, namely,
\begin{displaymath}
 G_z(t)=A_0 cos(\omega t + \varphi)exp({-\sigma ^2 t^2}) + A_1 exp(-\lambda t)+ C
\end{displaymath}
\noindent where $A_{0 }$ and $A_{1}$ are the initial asymmetries, $\omega$ is the precession frequency, $\phi$ is the phase, $\lambda$ and $\sigma$ are the depolarization rates, and C is the background. The temperature dependencies of these parameters are shown in Fig.~11. Fig.~11(a) clearly shows that at 12 K there is a loss of asymmetry in A$_{1 }$ to 1/3 the high temperature value. At the same temperature an increase in $A_{0}$ is observed together with an increasing frequency (see Fig.~11(b)), indicating the presence of a long range ordered state in PrRuSi$_{3}$ which contrasts the specific heat observation of no long  range magnetic ordering. The solid line in Fig.~11(b) is a guide to the eye and emphasizes the salient feature, namely that at the points of interest in the magnetization, T = 5 and 9 K, the frequency shows plateaus which indicates a spin-reorientation. Figures 11(c) and 11(d) show the temperature dependence of the muon depolarization rates. It is interesting to note that $\sigma$ continues to increase as the temperature is reduced, which is contrary to the expected temperature dependence for a static long range magnetically ordered system. This may suggest that there may be a degree of frustration in PrRuSi$_3$ in line with the magnetization and specific heat data. Furthermore, the support of the spin-glass behavior in a homogenous system arises from the well-defined crystal field excitations in this compound. As expected the temperature dependence of $\lambda$ shows a peak at the ordering temperature and also continues to rise as the temperature is reduced, again indicating slow down of spin dynamics.

\begin{figure}
\includegraphics[width=9.0cm, keepaspectratio]{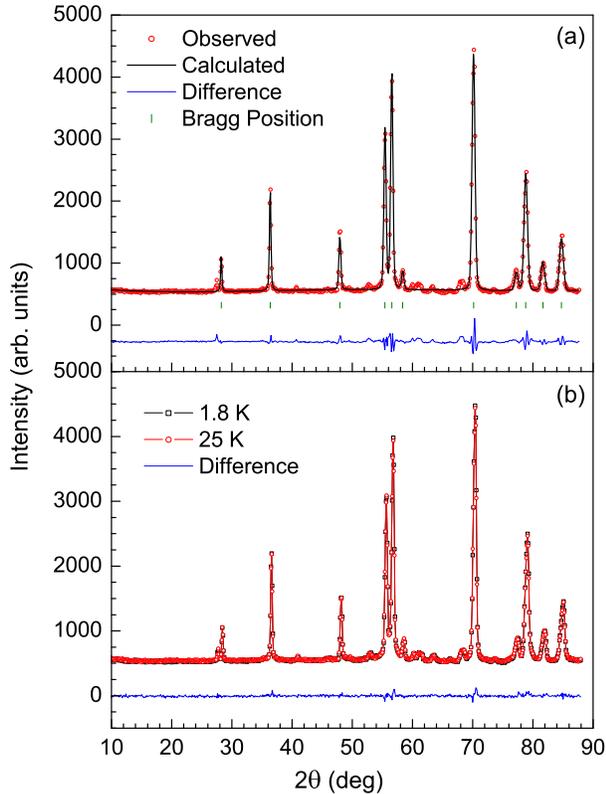}
\caption{\label{fig12} (colour online) (a) Neutron diffraction pattern of PrRuSi$_{3}$ recorded at 25~K. The solid line through the experimental points is the Rietveld refinement profile calculated for BaNiSn$_{3}$-type tetragonal (space group \textit{I4 mm}) structural model. The vertical bars indicate the Bragg positions. The lowermost curve represents the difference between the experimental data and calculated results. Neutron diffraction pattern of PrRuSi$_{3}$ recorded at 1.8 and 25~K together with their difference (b). No magnetic Bragg peaks are observed in the difference plot.}
\end{figure}

In order to explore the nature of long range order observed in $\mu$SR study we performed a low temperature neutron diffraction. The neutron diffraction patterns of PrRuSi$_{3}$ recorded at 1.8 and 25~K are shown in Fig.~12. A comparison of 1.8 and 25 K data (above and below the transition temperature) reveals no magnetic Bragg peak (Fig.~12(b)), the very small peaks observed in difference plot are due to the imperfect subtraction of data sets. This indicates that the long range ordered state moment of the Pr ion is below 0.2(1) $\mu_B$. A structural Rietveld refinement of 25 K neutron diffraction pattern (Fig.~12(a)) reveals single phase nature of sample with crystallographic parameters very close to what have been extracted from the room temperature X-ray diffraction pattern. At 25 K the lattice parameters are $a$ = 4.2162(5) {\AA} and $c$ = 9.9339(14) {\AA}, and that at 2 K are $a$ = 4.2161(5) {\AA} and $c$ = 9.9340(14).

\begin{figure}
\includegraphics[width=7.5cm, keepaspectratio]{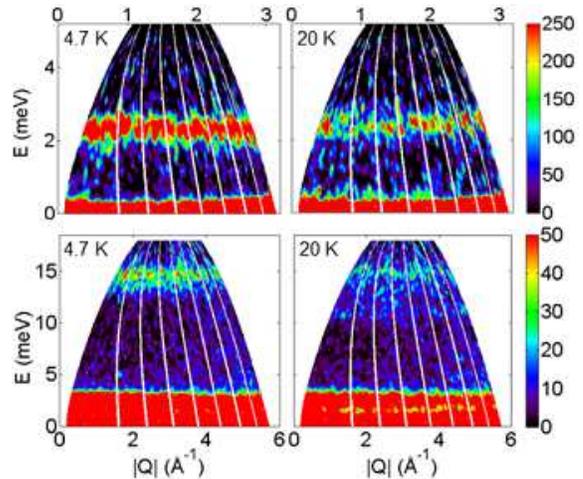}
\caption{\label{fig13} (colour online) Inelastic neutron scattering response, a color-coded map of the intensity, energy transfer ($E$) versus momentum transfer ($Q$) of PrRuSi$_{3}$ measured with the incident energy $E_{i}$ = 6 meV (top) and 20 meV (bottom) at 4.7 K and 20 K.}
\end{figure}

As discussed above the $\mu$SR study shows an evidence for the presence of long range magnetic order, however the specific heat and neutron diffraction results suggest that there is no long range magnetic ordering in PrRuSi$_{3}$, it seems that the moment is weak which is sensed by muons but not manifested in the specific heat and neutron diffraction measurements. A rough estimate of magnetic moment from $\mu$SR suggests a moment of $\sim$ 0.05 $\mu_B$ in PrRuSi$_3$. Further, considering that the ground state is a singlet, the observation of weak moment in $\mu$SR measurements suggest an induced magnetic moment behavior in PrRuSi$_{3}$ due to low lying crystal field excitations. Therefore in order to understand the origin of spin-glass and/or induced moment magnetism in PrRuSi$_{3}$ we have performed the inelastic neutron scattering (INS) study to investigate crystal field excitations and their energy scheme. Neutrons with incident energy $E_{i}$ = 6, 20 and 40 meV were used to record the INS spectra at 4.7, 20 and 50 K for scattering angles between 3$^{o}$ and 135$^{o}$. The spectra have been corrected for the background signal and the absolute normalized response (in units of mbr/sr/meV), using vanadium standard, is presented in Fig.~13 as a color-coded map of the intensity, energy transfer versus momentum transfer, for scattering from PrRuSi$_{3}$ measured with 6 meV and 20 meV incident neutron energies at 4.7 K and 20 K. The presence of sharp inelastic excitations near 2.4 meV and 14.7 meV are clearly observed in the INS spectra at 4.7 K. Further we observe that the peak position of both 2.4 meV and 14.7 meV excitations remain nearly invariant from 4.7 to 50 K, which suggest that these excitations are the result of the crystal field effects, indirectly reflecting an absence of spin wave contributions, which one would expect below the magnetic ordering temperature. Consistent with neutron diffraction data we did not find any magnetic Bragg peak in $Q$-cuts (integrated over the elastic line from $-$1 to 1 meV) at 4.7 K, which suggests that long range order if present in this compound has very small magnetic moment, which can not be detected in the present data. At 20 K we can see a weak intensity near 12.3 meV, which is due to excited transition from the ground state to excited state doublet (Fig.~13, right bottom). The $Q$-dependent integrated intensity between 2 and 3 meV at 4.7 K follows the square of Pr$^{3+}$ magnetic form factor [$F^{2}(Q)$] (Fig.~14), which suggests that the inelastic excitations result mainly from single-ion CEF-transitions. The $F^{2}(Q)$ behavior of the $Q$-dependent integrated intensity indicates a very small phonon contribution for $|Q|$ $<$ 3 which we have used for crystal field analysis. A similar behavior was also found for 14.7 meV peak. We have therefore not corrected the spectra for phonon contribution and we presumed that the presence of weak phonon part, if it is present would not affect our analysis for the low $|Q|$ data based on the CEF model.

\begin{figure}
\includegraphics[width=8.5cm, keepaspectratio]{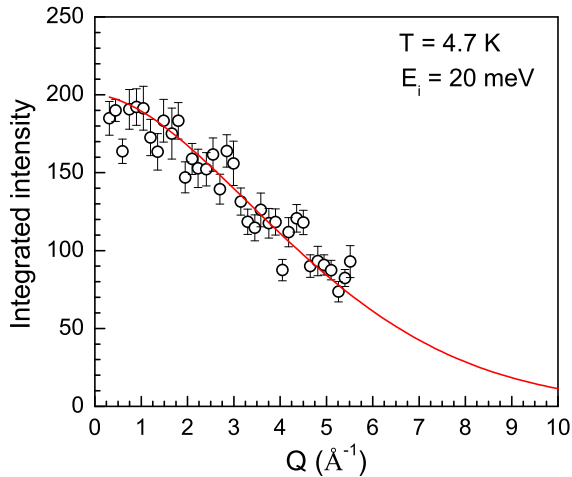}
\caption{\label{fig14} (colour online) The $Q$ dependence of total intensity integrated between 2 and 3 meV at 4.7 K for incident energy $E_{i}$ = 20 meV. The solid line represents the square of the Pr$^{3+}$ magnetic form factor, scaled to 200 at $Q$ = 0.}
\end{figure}

\begin{figure}
\includegraphics[width=7.5cm, keepaspectratio]{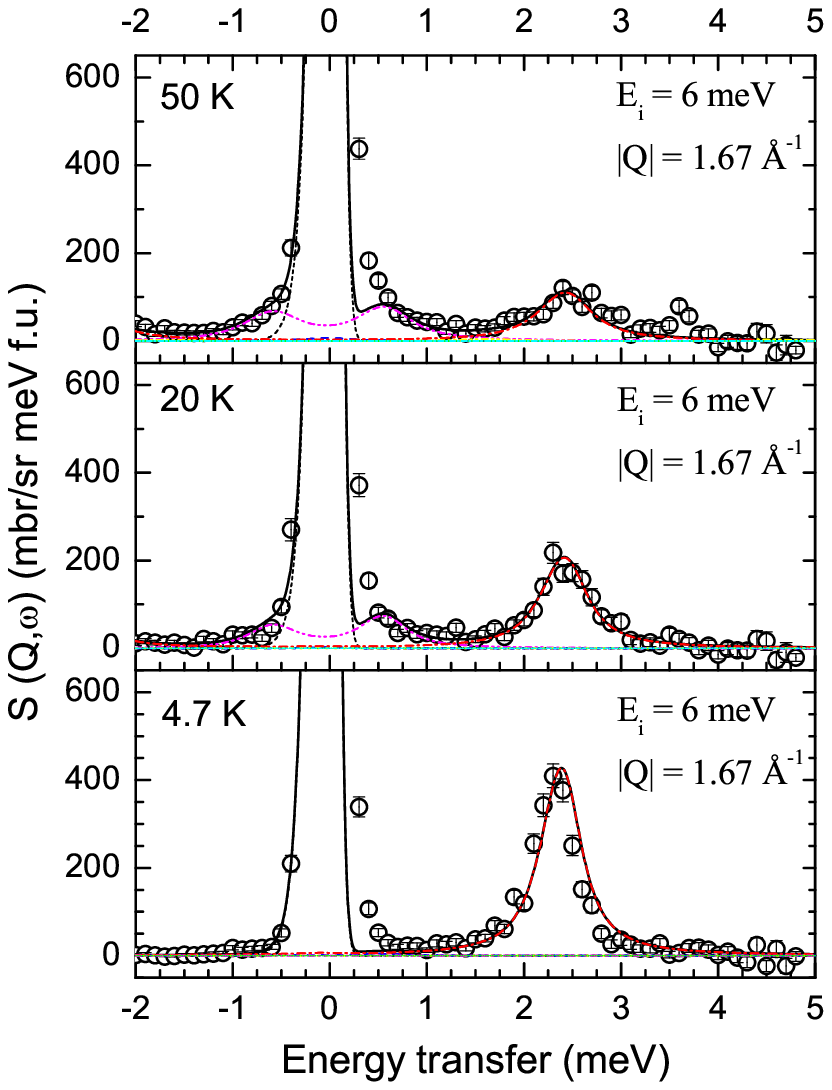}
\caption{\label{fig15} (colour online) The inelastic neutron scattering spectra of PrRuSi$_{3}$ measured with incident energy $E_{i}$ = 6 meV for momentum transfer $|Q|$ less than 3 {\AA}$^{-1}$ at different temperatures. The solid lines represent the crystal electric field fits. Different dotted, dashed, dashed-dotted lines correspond to the different components of the fit.}
\end{figure}

\begin{figure}
\includegraphics[width=7.5cm, keepaspectratio]{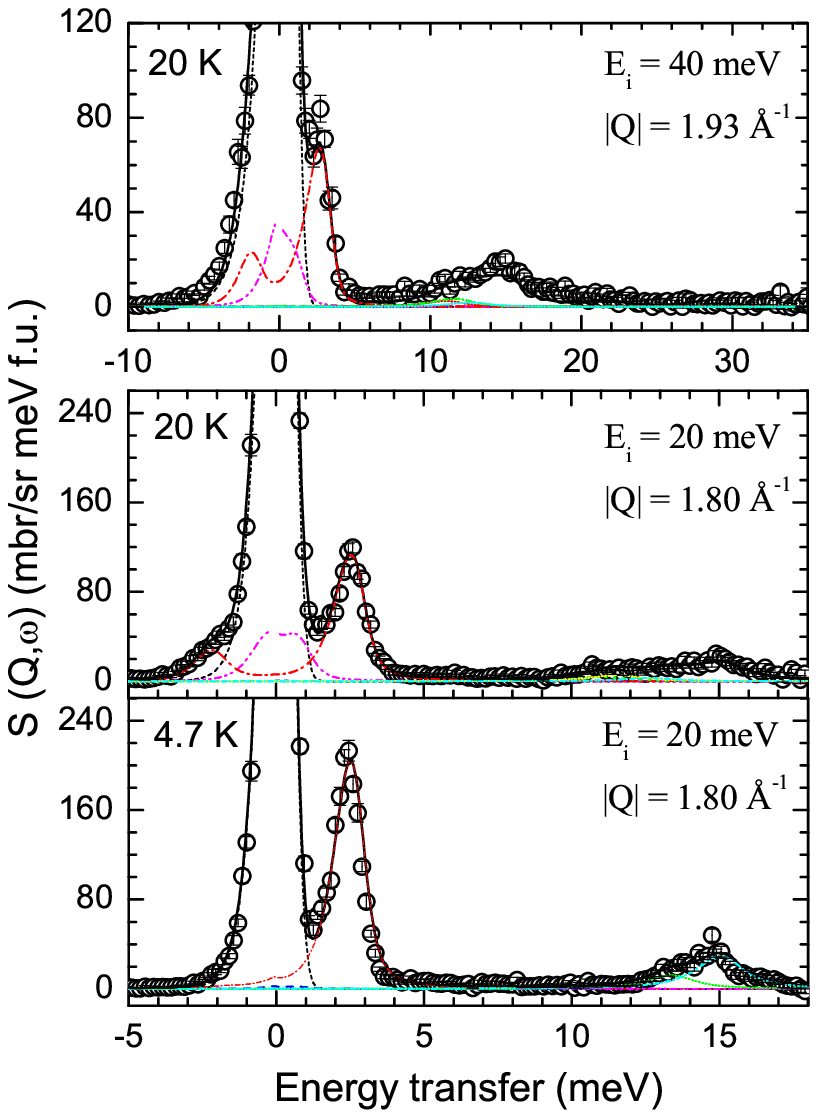}
\caption{\label{fig16} (colour online) The inelastic neutron scattering spectra of PrRuSi$_{3}$ measured with incident energy $E_{i}$ = 20 and 40 meV for momentum transfer $|Q|$ less than 3 {\AA}$^{-1}$ at different temperatures. The solid lines represent the crystal electric field fits. Different dotted, dashed, dashed-dotted lines correspond to the different components of the fit.}
\end{figure}

\begin{figure}
\includegraphics[width=8.5cm, keepaspectratio]{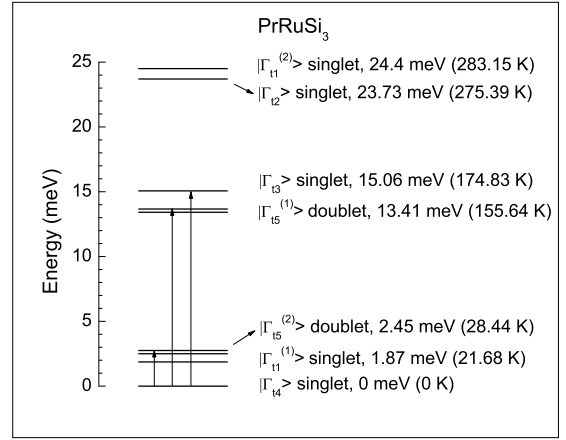}
\caption{\label{fig17} (colour online) Crystal electric field level scheme of the Pr$^{3+}$ ions in PrRuSi$_{3}$ deduced from the inelastic neutron scattering experiment. The transitions from the ground state to the excited states that contribute to the observed excitations are shown by arrows.}
\end{figure}

The INS spectra were analyzed further to obtain the full information about crystal field levels scheme. The crystal field Hamiltonian for the Pr ion having a tetragonal point group ($D_{4h}$) symmetry in PrRuSi$_{3}$ is given by
\[
H_{CEF} = B_{2}^{0}O_{2}^{0} + B_{4}^{0}O_{4}^{0} + B_{4}^{4}O_{4}^{4} + B_{6}^{0}O_{6}^{0} + B_{6}^{4}O_{6}^{4}
\]
\noindent where $O_{n}^{m}$ are the Stevens operator and $B_{n}^{m}$ are the phenomenological crystal field parameters that is determined from the experimental results of inelastic neutron scattering. The action of crystal electric field tends to remove the degeneracy of the 4\textit{f} ground multiplet. For tetragonal symmetry the nine fold-degenerate ground state of Pr$^{3+}$ ($J =4$) splits in five singlets ($\Gamma_{t1}^{(1)}$, $\Gamma_{t1}^{(2)}$, $\Gamma_{t2}$, $\Gamma_{t3}$, $\Gamma_{t4}$) and two doublets ($\Gamma_{t5}^{(2)}$, $\Gamma_{t5}^{(1)}$), the $\Gamma_{j}$'s are the irreducible representations of the point group.

The solid lines in Fig.~15 and Fig.~16 represent the fit to the CEF model for simultaneous refinement of all six data sets for 6, 20 and 40 meV incident energies (from 4.7, 20 and 50 K). The phenomenological crystal field parameters $B_{n}^{m}$ obtained from the best fit are $B_{2}^{0}$ = 0.4268 ($\pm$ 0.0025) meV, $B_{4}^{0 }$= 0.1031 ($\pm$ 0.0009) $\times$ 10$^{-2}$ meV, $B_{6}^{0}$ = 0.3151 ($\pm$ 0.0018) $\times$ 10$^{-4}$ meV, $B_{4}^{4}$ = -0.1996 ($\pm$ 0.0016)$ \times$ 10$^{-1}$ meV, and $B_{6}^{4}$ = 0.1563 ($\pm$ 0.0008) $\times$ 10$^{-2}$ meV. The crystal field level scheme corresponding to these $B_{n}^{m}$ parameters has a singlet ($\Gamma_{t4}$) ground state followed by a singlet ($\Gamma_{t1}^{(1)}$) at 1.87 meV, a doublet ($\Gamma_{t5}^{(2)}$) at 2.45 meV, another doublet ($\Gamma_{t5}^{(1)}$) at 13.41 meV, and three singlets $\Gamma_{t3}$, $\Gamma_{t2}$ and $\Gamma_{t1}^{(2)}$ respectively at 15.06 meV, 23.73 meV and 24.40 meV.  The crystal field level scheme of the Pr$^{3+}$ ions in PrRuSi$_3$ obtained this way is shown in Fig.~17. To compare, spin-glass system PrAu$_2$Si$_2$ also has a singlet $\Gamma_{t4}$ ground state, lying below a doublet first excited state $\Gamma_{t5}^{(2)}$ at 0.72 meV and second excited singlet $\Gamma_{t1}^{(1)}$ at 7.18 meV. \cite{30}

It is worth to mention here that though the first excited state is a singlet at 1.87 meV, there is no transition from the ground state to the first excited state, the matrix element is zero, the nonzero matrix elements for the transitions from ground state to the excited states are obtained only for the transitions shown by arrows in Fig.~17. Thus at temperatures close to the freezing temperature, which is well below 155 K (13.41 meV), PrRuSi$_3$ is effectively a two-level system. The specific heat calculated with the crystal field scheme presented in Fig.~17 is shown by solid line in Fig.~8. A good agreement between the experimentally observed specific heat data and the calculated one validates the obtained CEF level scheme.

\begin{figure}
\includegraphics[width=8.5cm, keepaspectratio]{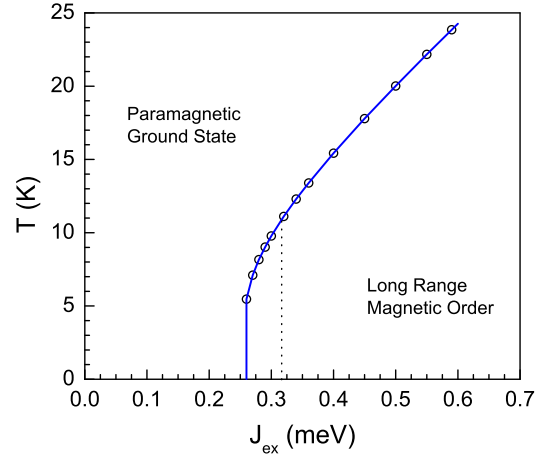}
\caption{\label{fig18} (colour online) The mean-field calculated $T_c$ as a function of the exchange energy, $J_{ex}$, for the singlet-doublet transition, $\Delta$ = 2.45 meV. The vertical dotted line shows $J_{ex}$ = 0.32 meV for PrRuSi$_{3}$, which gives ordering temperature $\sim$ 11.5 K that is very close to peak in the susceptibility at 9.8 K.}
\end{figure}

Now let us discuss the $\mu$SR finding of weak moment taking into consideration that the PrRuSi$_{3}$ has a singlet ground state as deduced from inelastic neutron scattering and specific heat measurements. For a singlet ground state system the existence of magnetic order critically depends on the ratio $J_{ex} / \Delta$, where $J_{ex}$ is the Heisenberg exchange interaction and $\Delta$ is the crystal field splitting energy between the ground state and the excited state coupled by the matrix element $\alpha$. \cite{31} Above a critical value of $J_{ex} / \Delta$ system undergoes a self-induced spontaneous ordering. The mean-field critical temperature below which self-induced moment forms spontaneously in a two-level system is given by \cite{16}
\[
T_{c} =  \Delta \left\{ ln \left[ \frac{J_{ex} \alpha^{2} + n \Delta}{J_{ex} \alpha^{2} - n \Delta}\right] \right\}^{-1}
\]
\noindent where $n$ is the degeneracy of the excited state and $\alpha$ is the the matrix element between the ground state singlet and excited state doublet. A plot of the $T_c$ calculated from this expression as a function of $J_{ex}$ for $n$ = 2, $\alpha^2$ = 9.58 and $\Delta$ = 2.45 meV is shown in Fig.~18. From Fig.~18 it follows that the exchange energy must be $J_{ex}$ = 0.30 meV for PrRuSi$_3$ to have an induced moment ordering below $T_c$ = 9.8 K. A rough estimate of $J_{ex}$ following the approach used to estimate the $J_{ex}$ for Pr in reference 29, i.e. using $\Delta \approx 2(g_J - 1) I_{sf}<J_z>$ and $J_{ex} = 0.85 I_{sf}$, yields $J_{ex}$ = 0.32 for PrRuSi$_3$ which is very close to the mean-field required value of $J_{ex}$ for an induced moment magnetic ordering below 9.8 K. Therefore this clearly suggests that PrRuSi$_{3}$ undergoes an induced moment ordering below 9.8 K. The value of $J_{ex}/\Delta$ = 0.122 for PrRuSi$_{3}$ is comparable but little higher than the corresponding $J_{ex}/\Delta$ = 0.085 for PrAu$_{2}$Si$_{2}$. \cite{16} Furthermore, as in the case of PrAu$_{2}$Si$_{2}$, if the exchange coupling is very close to the critical value required to induce the magnetic order, the dynamic fluctuations of crystal field levels can destabilize the induced moment long range magnetic order resulting in a frustrated ground state, i.e. spin-glass type behavior. The dynamic fluctuations of crystal field levels is the leading mechanism for spin-glass behavior in PrAu$_{2}$Si$_{2}$. \cite{16} For our system PrRuSi$_{3}$ it seems that the J$_{ex}$/$\Delta$ is higher than the critical value, resulting in an induced magnetic order as sensed by $\mu$SR, however the value of $J_{ex}/\Delta$ is not high enough to be unaffected by the dynamic fluctuations of crystal field levels and a frustrated ground state is realized together with induced moment magnetism. The two contradictory ground states, spin-glass and magnetic order, can be explained if the dynamic fluctuation time of the induced moment is smaller than the muon probing time window. In this case, the muon will observe the presence of an internal field, whereas the magnetization will observe the spin-glass behavior.

\section{Conclusions}

We have investigated the magnetic and transport properties of PrRuSi$_{3}$ using various experimental techniques. Our dc and ac magnetization studies reveal the signatures of non-canonical spin-glass behavior in this compound with the freezing temperature $T_{f}$ $\sim$ 9.8 K. The thermo-remnant magnetization in spin-glass state is found to relax with a very high mean relaxation time $\tau$ = 13357 s. The magnetic and thermal properties are strongly influenced by the crystal field effect that manifest as a broad anomaly in the specific heat and electric resistivity. The $\mu$SR data show the presence of long range magnetic ordering below 12 K with very small internal field at the muon site. The inelastic neutron scattering confirms a singlet CEF ground state and a possibility of the induced moment magnetism arising from the excited CEF doublet at 2.45 meV in the presence of strong exchange in PrRuSi$_{3}$. The CEF level scheme of PrRuSi$_{3}$ is very similar to that of PrAu$_{2}$Si$_{2}$ in which the spin-glass behavior arising from the dynamic fluctuations of the crystal field levels has been observed. This leads us to believe that the origin of spin-glass behavior in PrRuSi$_3$ lies in the dynamic fluctuations of crystal field levels. Considering the two contradictory ground states of PrRuSi$_{3}$, spin-glass behavior seen in the magnetization and specific heat study and long range ordering observed in the $\mu$SR study, we suspect that the time of dynamic fluctuations of the induced moments is smaller than the muon probing time window, therefore muon observes the presence of an internal field and hence the ordered state, whereas the magnetization observes the spin-glass behavior. Further, the value of the transition temperature of PrSuSi$_{3}$ increases with frequency of the probe, i.e., the ac susceptibility shows the peak at 10 K (at 10 KHz), while muon frequency in MHz shows the highest ordering temperature at 12 K. Thus the overall behavior of PrRuSi$_{3}$ leads us to characterize it as an induced moment spin-glass system, however further investigations preferably on single crystals are highly desired to understand better the mechanism of spin-glass behavior and induced moment magnetism in this well crystallographically ordered compound.

\acknowledgments

Authors VKA, DTA and ADH would like to acknowledge financial assistance from CMPC-STFC grant number CMPC-09108. We would like to thank Dr. Eugene Goremychkin for an interesting discussion.

\end{document}